\documentclass[aps,prd,twocolumn,groupedaddress,showpacs,amsmath,amssymb]{revtex4}
\usepackage{graphicx,epsf}
\usepackage[]{latexsym}
\newcommand{\be}{\begin{eqnarray}}
\newcommand{\ee}{\end{eqnarray}}

\newcommand{\expecl}[1]{\mbox{$\left\langle\,
            \strut\displaystyle{#1}\,\right\rangle$}}

\begin{document}
\title{Electromagnetic waves around dilatonic stars and
naked singularities}
\author{Roberto Casadio}
\email{casadio@bo.infn.it}
\affiliation{Dipartimento di Fisica, Universit\`a di Bologna, and
I.N.F.N., Sezione di Bologna, via Irnerio 46, 40126 Bologna, Italy}
\author{Sergio Fabi}
\email{fabi001@bama.ua.edu}
\author{Benjamin Harms}
\email{bharms@bama.ua.edu}
\affiliation{Department of Physics and Astronomy,
The University of Alabama,
Box 870324, Tuscaloosa, AL 35487-0324, USA}
\begin{abstract}
We study the propagation of classical electromagnetic waves on
the simplest four-dimensional spherically symmetric metric with
a dilaton background field.
Solutions to the relevant equations are obtained perturbatively
in a parameter which measures the strength of the dilaton field
(hence parameterizes the departure from Schwarzschild geometry).
The loss of energy from outgoing modes is estimated as a
back-scattering process against the dilaton background, which
would affect the luminosity of stars with a dilaton field.
The radiation emitted by a freely falling point-like source
on such a background is also studied by analytical and numerical
methods.
\end{abstract}
\maketitle
\section{Introduction}
\label{intro}
The occurrence of violations of the Equivalence Principle is a
subject of continuing investigation in the community of general
relativists, both theoretically and experimentally~\footnote{A
complete list of references for this issue would go beyond our
scope and necessity, and we just refer the reader to~\cite{will}
and References therein.}.
A possible source for such effects would be the existence of
a scalar component of gravity, which naturally emerges in string
theories and is usually referred to as the {\em dilaton\/}.
Solar system tests rule out such a possibility to a very high
precision in the weak field regime \cite{will}.
However, it cannot be excluded that the picture is different
in regions of strong gravitational fields, such as those that
can be found around very dense compact sources.
It is therefore interesting to study solutions of the
Einstein field equations with a scalar field.
\par
In this paper, we study the propagation of electro-magnetic
(EM) waves on the background of the Janis-Newman-Winicour (JNW)
solution \cite{jnw}.
The corresponding metric is spherically symmetric and
asymptotically flat, but the static dilaton field is not
trivial.
Although the JNW space-time is known to contain a naked
singularity, it is possible to make it physically useful by
suitably cutting the space outside the singular point and
then sewing a (spherically symmetric) regular inner patch
to complete the space-time manifold.
Provided this procedure is carried out consistently,
one can then describe compact astrophysical sources such as
stars with a dilaton field.
In this context, the JNW metric has recently attracted some
interest in relation to (strong) gravitational lensing effects
\cite{bozza}.
\par
When the source of EM radiation is placed at the center and is
also spherically symmetric, a convenient way of deriving the
relevant wave equations is to make use of the Newman-Penrose (NP)
formalism \cite{chandra}.
The exact equations thus obtained would however be exceedingly
complicated in the full-fledged JNW metric.
Therefore, in order to make some progress, we shall introduce a
parameter which is related to the asymptotic strength of the
dilaton field and can be smoothly perturbed away from the
Schwarzschild limit.
One then finds that the corresponding wave equations can be solved
analytically when the dilaton is sufficiently weak far away from
the source and expressions for the ingoing and outgoing modes
will be obtained in this approximation.
As a straightforward application of this analysis, we shall
consider the relative darkening of radiation emitted by a star
placed at the center of the system with respect to the vacuum
Schwarzschild case.
The main result is that such an effect turns out to be inversely
proportional to the star radius and to depend on the emitted energy
(but not on the wave frequency, contrary to what was found for a
rotating black hole \cite{knd0} in Ref.~\cite{knd2}).
\par
A more complicated problem is to obtain the spectrum of the
radiation emitted by a freely falling, electrically charged,
point-like source in the JNW background.
The wave equations now contain the energy-momentum density of
the falling particle and must be solved numerically for various
values of the perturbing parameter in order to compare the emitted
power to the analogous case in Schwarzschild.
A preliminary analytical analysis will show that some effect is
already expected at large distance from the center, however most
radiation would be emitted near the singularity where tidal forces
on the EM field of the falling particle become stronger.
Since naked singularities seem to be possible intermediate stages
in the gravitational collapse of regular matter (see, e.g.
\cite{NSreview} and References therein), one can view the falling
source as a component of the forming collapsed object.
Our results may then contribute to the understanding of the
(in)stability of naked singularities. The intensity of the radiation
shows an interference-like effect which would provide observers with
a distinctive signature for the presence of a dilatonic component
in strong gravitational fields.
\par
In the following Section, we shall review the JNW solution and
express it in terms of the parameter which allows the Schwarzschild limit
to be approached in a smooth manner. This property will be used later to
define a quasi-Schwarzschild regime, particularly suited for
analyzing (asymptotically) small deviations away from
Schwarzschild. We then show in Section~\ref{waves} how to
study electromagnetic waves on such a background using the NP
formalism \cite{chandra}. In Section~\ref{free} we solve the
Maxwell equations using Green function techniques. We finally
summarize and comment our results in Section~\ref{conc}.
\par
We use units with $G=c=1$ and follow the conventions of
Ref.~\cite{chandra}.
\section{JNW solution and Schwarzschild limit}
\label{background}
The JNW metric is a static, spherically symmetric and electrically
neutral solution of the field equations obtained by varying the
action
\be
S&=&{1\over 16}\,\int d^4x\,\sqrt{-g}\,\left[R
-{1\over 2}\,\left(\nabla\Phi\right)^2\right]
\nonumber
\\
&&-\int d^4x\,\sqrt{-g}\,e^{-q\,\Phi}\,F^2
\ ,
\ee
in which $R$ is the scalar curvature of the metric $g_{\mu\nu}$,
whose determinant is denoted by $g$, $F$ is the EM field strength
and note that the dilaton $\Phi$ is included inside the gravitational
Lagrangian.
Further, the coupling constant $q>0$ between the EM
field and the dilaton has also been exponentiated.
\par
For the purpose of studying linear perturbations in the
NP formalism~\cite{chandra}, it is convenient to write
the JNW metric \cite{jnw} as
\be
ds^2={\Delta\over\rho^2}\,dt^2-{\rho^2\over\Delta}\,dr^2
-\rho^2\,d\Omega^2
\ ,
\label{metric}
\ee
where $d\Omega^2=d\theta^2+\sin^2\theta\,d\phi^2$,
\be
\begin{array}{l}
\rho^2=\left(r-r_-\right)^{1-a}\,\left(r-r_+\right)^{1+a}
\\
\\
\Delta=\left(r-r_-\right)\,\left(r-r_+\right)
\ ,
\end{array}
\ee
and the dilaton field is then
\be
\Phi=\sqrt{1-a^2}\,\ln\left({r-r_-\over r-r_+}\right)
\ .
\label{dilaton}
\ee
\par
Note that, with respect to the original Ref.~\cite{jnw},
$a=1/\mu$ and we have introduced new parameters
\be
\begin{array}{l}
r_-=+\strut\displaystyle{r_0\over 2}\,{1-a\over a}
\\
\\
r_+=-\strut\displaystyle{r_0\over 2}\,{1+a\over a}
\ .
\end{array}
\label{para}
\ee
Although the JNW metric contains only two independent
parameters \cite{jnw}, the Schwarzschild limit is more
conveniently obtained by treating $a$, $r_-$ and $r_+$
as independent (we just assume $r_-<r_+$) and separately taking
the limits
\be
r_-\to 0
\ \ \ \ {\rm and}\ \ \
a\to -1
\ .
\label{schw1}
\ee
The Schwarzschild horizon is then given by $r_h\equiv 2\,M=r_+$
\footnote{One can equivalently assume $r_+<r_-$ and then take
$r_+\to 0$ and $a\to +1$, in which case $r_h=r_-$.
Note that such limits could not be taken in the original
parameters $r_0$ and $\mu=1/a$ of Ref.~\cite{jnw}, since
$r_0=-(r_++r_-)>0$ and $0<a=(r_++r_-)/(r_+-r_-)<1$ are not
compatible with Eq.~(\ref{schw1}), or the alternate limits in
this footnote.
\label{f1}}.
\par
The relevant dilatonic quantity is the spatial derivative
\be
\partial_r\Phi=
{\sqrt{1-a^2}\,(r_--r_+)\over(r-r_-)\,(r-r_+)}
\ ,
\ee
which determines the energy-momentum tensor
\be
T_{\mu\nu}=
\delta^r_\mu\,\delta^r_\nu\,\left(\partial_r\Phi\right)^2
-{1\over 2}\,g_{\mu\nu}\,g^{rr}\,\left(\partial_r\Phi\right)^2
\ .
\label{Tmn}
\ee
In particular, the non-vanishing components diverge at
$r=r_+$,
\be
T^t_{\ t}\sim
T^r_{\ r}\sim
T^\theta_{\ \theta}\sim
T^\phi_{\ \phi}
\sim{1-a^2\over (r-r_+)^{2+a}}
\ .
\ee
and $r_+$ is therefore a naked real singularity.
Further, since $\rho(r_+)=0$, such a singularity is point-like
and values of $r<r_+$ can be considered as unphysical
\footnote{The components of the energy-momentum tensor in
Eq.~(\ref{Tmn}) also diverge at $r=r_-$.
However, since we have assumed $r_-<r_+$, there is no
physical point identified by $r=r_-$ in the space-time
of interest to us (i.e., $r>r_+$).}.
Let us finally note that the energy-momentum tensor properly
vanishes in the limit $a\to\pm 1$ (corresponding to the
Schwarzschild vacuum).
\par
The convenient feature of the limit in Eq.~(\ref{schw1}) is that
it is smooth in $r_-$ and $a$.
This allows us to introduce a quasi-Schwarzschild regime
[corresponding to a ``small'' deviation away from the limit
(\ref{schw1})] for the region outside the ``horizon'' ($0<r_h<r$).
It is defined by
\be
r_-=0
\ \ \ \ {\rm and}\ \ \
a=-1+2\,x^2
\ ,
\label{qschw}
\ee
with $0\le x\ll 1$.
In particular, for small $x$, one finds that the metric (\ref{metric})
reduces to Schwarzschild plus corrections of order $x^2$,
\be
ds^2&\simeq&
-\left[1-(1-2\,x^2)\,{r_h\over r}-x^2\,{r_h^2\over r^2}\right]\,dt^2
\nonumber
\\
&&+\left[1+(1-2\,x^2)\,{r_h\over r}\right]\,dr^2
+r^2\,d\Omega^2
\ ,
\ee
where $r_h\equiv r_+$.
The ADM mass for the above metric is given by the usual Schwarzschild
relation
\be
M={r_h\over 2}
\ ,
\label{adm}
\ee
and the relevant PPN parameters \cite{will} by~\footnote{For a general
post-Newtonian analysis of the JNW metric, see Ref.~\cite{finelli}.}
\be
\begin{array}{l}
\alpha=\gamma=1-2\,x^2
\\
\\
\beta=1-6\,x^2
\ .
\end{array}
\label{ppn}
\ee
In the same approximation, the derivative of the dilaton becomes
\be
\partial_r\Phi\simeq-{2\,x\over r-r_h}\,{r_h\over r}
\sim -2\,x\,{r_h\over r^2}
\ ,
\label{phiR}
\ee
from which one can see that $x$ is a measure of the dilaton strength
for large $r$ (note that $\partial_r\Phi$ diverges for $r\to r_h$
for all values of $x>0$).
\par
The Eqs.~(\ref{ppn}) yield a quantitative way of estimating how
small the deviations from the Schwarzschild solution are in the
weak field zone ($r\gg r_h$).
From solar test measurements one can therefore read off the bound
$x^2\lesssim 10^{-5}$ \cite{will}, which can then be substituted
into Eq.~(\ref{phiR}) to obtain the upper limit of the admissible
dilaton strength.
Since no direct measurement in the strong field regime
is available, the strength of the dilaton field cannot be bounded
for $r\sim r_h$.
\section{Dilatonic stars}
\label{waves}
As we mentioned in the Introduction, the first case we consider is
that of a spherically symmetric source for the EM waves placed at
the center, so that it can also be identified with the source of the
JNW gravitational and dilaton fields.
In order to study Maxwell's equations in this context, we shall take
advantage of the spherical symmetry by employing the NP formalism
\cite{chandra}.
\subsection{Maxwell waves in NP formalism}
We first choose the following contravariant components of the NP
tetrad,
\be
&&
l^i=\left[
\begin{array}{cccc}
{\rho^2\over\Delta}\ ,
& 1\ ,
& 0\ , & 0
\end{array}\right]
\nonumber
\\
&&
n^i=\left[
\begin{array}{cccc}
{1\over 2}\ ,
& -{\Delta\over 2\,\rho^2}\ ,
& 0\ , & 0
\end{array}\right]
\label{tetrad}
\\
&&
m^i=\left[
\begin{array}{cccc}
0\ , & 0\ ,
&{1\over\sqrt{2\,\rho^2}}\ ,
&{i\,{\rm cosec}(\theta)\over\sqrt{2\,\rho^2}}
\end{array}\right]
\ .
\nonumber
\ee
The Maxwell field is now completely described by the three complex
scalars
\be
\phi_0&\equiv&F_{ij}\,l^i\,m^j
\nonumber
\\
&=&{\rho^3\over\sqrt{2}\,\Delta}\,\left[
E^\theta+B^\varphi+i\,\sin(\theta)\,\left(E^\varphi-B^\theta\right)
\right]
\nonumber
\\
\phi_1&\equiv&{1\over 2}\,F_{ij}\,\left(l^i\,n^j+m^{i*}\,m^j\right)
\nonumber
\\
&=&{\rho^4\,E^r\over2\,\Delta^2}\,\left(1+{\Delta^2\over 4\,\rho^4}\right)
\label{phiEB}
\\
\nonumber
\\
\phi_2&\equiv&F_{ij}\,m^{i*}\,n^j
\nonumber
\\
&=&-{\rho\over2\,\sqrt{2}}\,\left[
E^\theta-B^\varphi+i\,\sin(\theta)\,\left(E^\varphi+B^\theta\right)
\right]
\nonumber
\ ,
\ee
where $E^i$ and $B^i$ are the usual electric and magnetic components.
The $\phi_i$ must satisfy the equations \cite{knd}
\begin{subequations}
\be
(\hat D-2\,\tilde\rho)\,\phi_1
-(\hat\delta^*+\tilde\pi-2\,\tilde\alpha)\,\phi_0
+\tilde\kappa\,\phi_2
=J_1
\label{max1}
\ee
\be
(\hat\delta-2\,\tilde\tau)\,\phi_1
-(\hat\Delta+\tilde\mu-2\,\tilde\gamma)\,\phi_0
+\tilde\sigma\,\phi_2=
J_3
\label{max2}
\ee
\be
(\hat D-\tilde\rho+2\,\tilde\epsilon)\,\phi_2
-(\hat\delta^*+2\,\tilde\pi)\,\phi_1
+\tilde\lambda\,\phi_0=
J_4
\label{max3}
\ee
\be
(\hat\delta-\tilde\tau+2\,\tilde\beta)\,\phi_2
-(\hat\Delta+2\,\tilde\mu)\,\phi_1
+\tilde\nu\,\phi_0=
J_2
\ ,
\label{max4}
\ee
\end{subequations}
with the dilatonic currents
\begin{subequations}
\be
J_1=
{q\over 2}\,\left[(\phi_1+\phi_1^*)\,\hat D
-\phi_0\,\hat\delta^*-\phi_0^*\,\hat\delta\right]\Phi
\label{J1}
\ee
\be
J_2=
{q\over 2}\,\left[(\phi_1+\phi_1^*)\,\hat\Delta
-\phi_2\,\hat\delta-\phi_2^*\,\hat\delta^*\right]\Phi
\label{J2}
\ee
\be
J_3=
{q\over 2}\,\left[(\phi_1-\phi_1^*)\,\hat\delta
-\phi_0\,\hat\Delta+\phi_2^*\,\hat D\right]\Phi
\label{J3}
\ee
\be
J_4=
{q\over 2}\,\left[(\phi_1-\phi_1^*)\,\hat\delta^*
+\phi_0^*\,\hat\Delta-\phi_2\,\hat D\right]\Phi
\ .
\label{J4}
\ee
\end{subequations}
In the above, tilded quantities are spin coefficients, and hatted
quantities represent differential operators.
In particular, the non-vanishing spin coefficients for the
JNW metric (\ref{metric}) and tetrad (\ref{tetrad}) are given by
\be
\tilde\rho&=&-\strut\displaystyle{2\,r+r_+\,(a-1)-r_-\,(a+1)
\over 2\,(r-r_-)\,(r-r_+)}
\nonumber
\\
\tilde\mu&=&\strut\displaystyle{2\,r+r_+\,(a-1)-r_-\,(a+1)
\over 4\,(r-r_-)^{1-a}\,(r-r_+)^{1+a}}
\nonumber
\\
\tilde\alpha&=&-\tilde\beta
\\
&=&
-\strut\displaystyle{\sqrt{2}\,{\rm cotan}(\theta)
\over 4\,(r-r_-)^{1-a\over 2}\,(r-r_+)^{1+a\over 2}}
\nonumber
\\
\tilde\gamma&=&\strut\displaystyle{a\,(r_--r_+)
\over 4\,(r-r_-)^{1-a}\,(r-r_+)^{1+a}}
\ .
\nonumber
\ee
Further, the time dependence of the EM waves is taken
to be $\phi_i\sim \exp\left(+i\,\omega\,t\right)$, so that the
differential operators appearing in the Maxwell equations become
\be
\begin{array}{l}
\hat D=\partial_r+i\,\strut\displaystyle{\rho^2\over\Delta}\,\omega
\\
\\
\hat\Delta=-\strut\displaystyle{\Delta\over 2\,\rho^2}\,
\left(\partial_r
-i\,{\rho^2\over\Delta}\,\omega\right)
\\
\\
\hat\delta=\strut\displaystyle{1\over \sqrt{2\,\rho^2}}\,
\left[\partial_\theta-m\,{\rm cosec}(\theta)\right]
\\
\\
\hat\delta^*=\strut\displaystyle{1\over \sqrt{2\,\rho^2}}\,
\left[\partial_\theta+m\,{\rm cosec}(\theta)\right]
\ .
\end{array}
\ee
\par
Upon substituting the above expressions into
Eqs.~(\ref{max1})-(\ref{max4}),
one obtains the following four equations~\footnote{Note that
$\Phi=\Phi(r)$ does not depend on the angular variables,
so that $\hat\delta\Phi=-\hat\delta^*\Phi=0$.}
\begin{widetext}
\begin{subequations}
\be
&&\!\!\!\!\!\!\!
\left[\partial_r+
{2\,r+r_+\,(a-1)-r_-\,(a+1)\over (r-r_-)\,(r-r_+)}
+i\left({r-r_+\over r-r_-}\right)^a\omega\right]
\phi_1
-{\partial_\theta
+{\rm cosec}(\theta)\,\left[m+\cos(\theta)\right]\over
\sqrt{2}\,(r-r_-)^{1-a\over 2}\,(r-r_+)^{1+a\over 2}}
\,\phi_0
\nonumber
\\
&&
\ \ \
={q\over 2}\,\left(\phi_1+\phi_1^*\right)\,\partial_r\Phi
\label{em1}
\ee
\be
&&\!\!\!\!\!\!\!
\left[\partial_r
+{2\,r-r_+\,(a+1)+r_-\,(a-1)\over 2\,(r-r_-)\,(r-r_+)}
-i\left({r-r_+\over r-r_-}\right)^a\omega\right]
\phi_0
+\sqrt{2}\,{\partial_\theta-m\,{\rm cosec}(\theta)\over
(r-r_-)^{1+a\over 2}\,(r-r_+)^{1-a\over 2}}\,
\phi_1
\nonumber
\\
&&
\ \ \
={q\over 2}\,\left[
\phi_0+2\,\left({r-r_+\over r-r_-}\right)^a\,\phi_2^*\right]\,
\partial_r\Phi
\label{em2}
\ee
\be
&&\!\!\!\!\!\!\!
\left[\partial_r+
{2\,r+r_+\,(a-1)-r_-\,(a+1)\over2\,(r-r_-)\,(r-r_+)}
+i\left({r-r_+\over r-r_-}\right)^a\omega
\right]
\phi_2
-{\partial_\theta+m\,{\rm cosec}(\theta)\over
\sqrt{2}\,(r-r_-)^{1-a\over 2}\,(r-r_+)^{1+a\over 2}}\,
\phi_1
\nonumber
\\
&&
\ \ \
=-{q\over 2}\,\left[
{\phi_0^*\over 2}\,\left({r-r_+\over r-r_-}\right)^a+\phi_2\right]
\partial_r\Phi
\label{em3}
\ee
\be
&&\!\!\!\!\!\!\!
\left[\partial_r+
{2\,r+r_+\,(a-1)-r_-\,(a+1)\over (r-r_-)\,(r-r_+)}
-i\left({r-r_+\over r-r_-}\right)^a\omega
\right]\phi_1
+\sqrt{2}\,{\partial_\theta
-{\rm cosec}(\theta)\,\left[m-\cos(\theta)\right]
\over (r-r_-)^{1+a\over 2}\,(r-r_+)^{1-a\over 2}}\,
\phi_2
\nonumber
\\
&&
\ \ \
=-{q\over 2}\,\left(\phi_1+\phi_1^*\right)\,\partial_r\Phi
\label{em4}
\ ,
\ee
\end{subequations}
\end{widetext}
which we now proceed to study in a suitable limit.
\subsection{Quasi-Schwarzschild solutions}
\label{quasiS}
We are interested in deviations from Schwarzschild induced
by a background dilaton (\ref{dilaton}) which is experimentally
unobservable in the weak field regime.
We will therefore look for solutions to Eqs.~(\ref{em1})-(\ref{em4})
in the quasi-Schwarzschild regime defined by Eq.~(\ref{qschw}),
bearing in mind that the parameter $x$ is constrained by solar
system tests via Eqs.~(\ref{ppn}) as we mentioned previously.
\par
Upon expanding in $x$, Eq.~(\ref{em1}) becomes
\begin{widetext}
\be
&&
\left[\partial_r
+{2\over r}\left(1+{x^2 r_h\over r-r_h}\right)
+{i\,\omega\,r\over r-r_h}
\left(1+2\,x^2 \ln{r-r_h\over r}\right)
\right]\phi_1
-\left(1-x^2\,\ln{r-r_h\over r}\right)
{\partial_\theta+{\rm cosec}(\theta)\,\left[m+\cos(\theta)\right]
\over\sqrt{2}\,r}\,\phi_0
\nonumber
\\
&&
\simeq
-q\,x\,{r_h\,\sqrt{1-x^2}\over r\,(r-r_h)}\,
\left(\phi_1+\phi_1^*\right)
\ ,
\ee
\end{widetext}
and analogous expressions are obtained for Eqs.~(\ref{em2})-(\ref{em4}).
Note that the currents on the right hand sides contain terms proportional
to $x$ which are not on the left hand sides.
One can then discard all terms of order $x^2$ (and higher) and finally
obtains
\begin{widetext}
\begin{subequations}
\be
\!\!\!\!\!\!\!\!\!\!\!\!\!\!\!\!\!\!\!\!\!\!\!\!\!\!\!\!\!\!\!\!\!\!\!\!\!
\!\!\!\!
\left(\partial_r+{2\over r}+{i\,\omega\,r\over r-r_h}\right)\phi_1
-{\partial_\theta+{\rm cosec}(\theta)\,\left[m+\cos(\theta)\right]
\over\sqrt{2}\,r}\,\phi_0
=-q\,x\,{r_h\,\left(\phi_1+\phi_1^*\right)\over r\,(r-r_h)}
\label{em11}
\ee
\be
\!\!\!\!\!\!\!\!\!\!\!\!\!\!
\left(\partial_r+{1-i\,\omega\,r\over r-r_h}\right)\phi_0
+\sqrt{2}\,\ln\left(1-{r_h\over r}\right)\,
{\partial_\theta-m\,{\rm cosec}(\theta)\over r-r_h}
\,\phi_1
=-{q\,x\,r_h\over r-r_h}\,
\left({\phi_0\over r}+{2\,\phi_2^*\over r-r_h}\right)
\label{em21}
\ee
\be
\!\!\!\!\!\!\!\!\!\!\!\!\!\!\!\!\!\!\!\!\!\!\!\!\!\!\!\!\!\!\!\!\!\!\!\!\!
\!\!\!\!\!\!\!\!\!\!\!\!\!\!\!\!\!\!\!
\left(\partial_r+{1\over r}+{i\,\omega\,r\over r-r_h}
\right)\phi_2
-{\partial_\theta+m\,{\rm cosec}(\theta)\over\sqrt{2}\,r}\,\phi_1
=q\,x\,{r_h\over r}\,
\left({\phi_0^*\over 2\,r}+{\phi_2\over r-r_h}\right)
\label{em31}
\ee
\be
\left(\partial_r+{2\over r}-{i\,\omega\,r\over r-r_h}
\right)\phi_1
+\sqrt{2}\,\ln\left(1-{r_h\over r}\right)
{\partial_\theta-{\rm cosec}(\theta)\,
\left[m-\cos(\theta)\right]\over r-r_h}
\,\phi_2
=q\,x\,{r_h\,\left(\phi_1+\phi_1^*\right)\over r\,(r-r_h)}
\label{em41}
\ ,
\ee
\end{subequations}
\end{widetext}
where the left hand sides equated to zero are just the wave
equations for the EM waves in Schwarzschild.
\par
It is now possible to expand the $\phi_i$'s in powers of $x$,
\be
\phi_i=\phi_i^{(0)}+x\,\phi_i^{(1)}
\ ,
\ee
where the $\phi_i^{(0)}$'s solve the above
Eqs.~(\ref{em11})-(\ref{em41}) with vanishing right hand sides.
Hence the $\phi_i^{(1)}$'s must solve Eqs.~(\ref{em11})-(\ref{em41})
with $x=1$.
\par
In the phantom gauge ($\phi_0=\phi_2=0$ \cite{chandra}),
$\phi_1^{(1)}$ should simultaneously satisfy the four equations
\begin{subequations}
\be
\left(\partial_r+{2\over r}\pm{i\,\omega\,r\over r-r_h}
\right)\phi_1^{(1)}
=\mp q\,{r_h\,\left(\phi_1^{(1)}+\phi_1^{(1)*}\right)\over r\,(r-r_h)}
\label{phantom1}
\ee
\be
\!\!\!\!\!\!\!\!\!\!\!\!\!\!\!\!\!\!
\!\!\!\!\!\!\!\!\!\!\!\!\!\!\!\!\!\!
\!\!\!\!\!\!\!\!\!\!\!\!\!\!\!\!\!\!
\!\!\!\!\!\!\!
\left[\partial_\theta\pm{m\over\sin(\theta)}\right]\phi_1^{(1)}
=0
\ .
\label{phantom2}
\ee
Eqs.~(\ref{phantom2}) imply that $\phi_1^{(1)}=\phi_1^{(1)}(r)$.
Further, Eqs.~(\ref{phantom1}) can be written as
\be
\begin{array}{l}
\strut\displaystyle{\left(\partial_r+{2\over r}\right)\phi_1^{(1)}}
=0
\\
\\
\ \
i\,\omega\,\phi_1^{(1)}=
-2\,q\,\strut\displaystyle{r_h\over r}\,{\rm Re}
\left(\phi_1^{(1)}\right)
\ ,
\end{array}
\ee
\end{subequations}
and the second condition leads to $\phi_1^{(1)}=0$.
Since the phantom gauge usually holds for the background solution,
this is in agreement with the fact that there is no background EM
field in the JNW solution.
\par
For EM waves it is possible to set $\phi_1^{(1)}=0$
\footnote{From Eq.~(\ref{phiEB}), one finds that $\phi_1=0$ implies
$E^r=0$ and there is no radial oscillation in the EM field
(as is expected for spherically symmetric waves).}
and define
\be
\begin{array}{l}
W_0\equiv\phi_0^{(1)}
\\
\\
W_2\equiv -2\,\phi_2^{(1)*}
\ ,
\end{array}
\ee
which must then solve the equations~\footnote{Eq.~(\ref{em32})
follows from the complex conjugate of Eq.~(\ref{em31}).}
\begin{subequations}
\be
\!\!\!\!\!\!\!\!\!\!\!\!\!\!\!\!\!\!
\!\!\!\!\!\!\!\!\!\!\!\!\!\!\!\!\!\!
\!\!\!\!\!\!\!\!\!\!\!\!\!\!\!\!\!\!
\left[\partial_\theta+{m+\cos(\theta)\over\sin(\theta)}
\right]W_0
=0
\label{em12}
\ee
\be
\left(\partial_r+{1-i\,\omega\,r\over r-r_h}\right)W_0
={q\,r_h\over r-r_h}\,
\left({W_2\over r-r_h}-{W_0\over r}\right)
\label{em22}
\ee
\be
\left(\partial_r+{1\over r}-{i\,\omega\,r\over r-r_h}\right)W_2
={q\,r_h\over r}\,
\left({W_2\over r-r_h}-{W_0\over r}\right)
\label{em32}
\ee
\be
\!\!\!\!\!\!\!\!\!\!\!\!\!\!\!\!\!\!
\!\!\!\!\!\!\!\!\!\!\!\!\!\!\!\!\!\!
\!\!\!\!\!\!\!\!\!\!\!\!\!\!\!\!\!
\left[\partial_\theta-{m-\cos(\theta)\over\sin(\theta)}
\right]W_2^*
=0
\label{em42}
\ .
\ee
\end{subequations}
Note that the angular parts of $W_0\equiv S_0(\theta,\varphi)\,R_0(r)$
and $W_2\equiv S_2(\theta,\varphi)\,R_2(r)$ satisfy the independent
Eqs.~(\ref{em12}) and (\ref{em42}).
They are thus spherical harmonics with $S_0=S_2=Y_l^m(\theta,\varphi)$,
the same as for $q=0$.
Consequently, the remaining Eqs.~(\ref{em22}) and (\ref{em32})
become purely radial equations for the $R_i(r)$'s,
\be
\begin{array}{l}
R_0(r)=e^{i\,\sigma(r)}\,X(r)
\\
\\
R_2(r)=e^{i\,\sigma(r)}\,Y(r)
\ .
\end{array}
\ee
The real phase $\sigma$ is the same for both fields and satisfies
the purely Schwarzschild equation
\be
{d\sigma\over dr}={\omega\,r\,\sigma\over r-r_h}
\ .
\label{k}
\ee
Hence, $\phi_0\sim R_0$ represents the ingoing modes while
$\phi_2\sim R_2^*$ stands for the outgoing modes.
The presence of terms (proportional to $q$) which mix ingoing
with outgoing modes immediately implies that outgoing radiation
emitted by the central source would partly back-scatter
into ingoing waves.
Thus the expected dilaton effect is to make the source less bright
(see the next Subsection for more quantitative estimates).
\par
The general solutions to the homogeneous equations are given
by
\be
\begin{array}{l}
\phi_0^{(0)}=A\,\strut\displaystyle{e^{+i\,\sigma(r)}\over r-r_h}
\,Y_l^m(\theta,\varphi)
\\
\\
\phi_2^{(0)}=B\,\strut\displaystyle{e^{-i\,\sigma(r)}\over r}
\,Y_l^{-m}(\theta,\varphi)
\ ,
\end{array}
\label{homo}
\ee
where $\sigma$ solves Eq.~(\ref{k}) and the constants $A$ and
$B$ must be determined from the initial conditions.
\par
The functions $X$ and $Y$ must solve
\begin{subequations}
\be
\left(\partial_r+{1\over r-r_h}\right)X
={q\,r_h\over r-r_h}\,
\left({Y\over r-r_h}-{X\over r}\right)
\label{em23}
\ee
\be
\!\!\!\!\!\!\!\!\!\!\!\!\!\!\!
\left(\partial_r+{1\over r}\right)Y
={q\,r_h\over r}\,
\left({Y\over r-r_h}-{X\over r}\right)
\ .
\label{em33}
\ee
\end{subequations}
Since $E^2\sim B^2\sim 1/r^2$ for large $r$, admissible solutions
to Eqs.~(\ref{em23}) and (\ref{em33}) should be of order $1/r$ or
higher at large $r$, as occurs to the $\phi_i^{(0)}$'s.
In fact, the particular solutions are given by
\be
\begin{array}{l}
X=\strut\displaystyle{C\over r-r_h}\,
\left[q\,\ln\left(1-{r_h\over r}\right)-1\right]
\\
\\
Y=q\,\strut\displaystyle{C\over r}\,\ln\left(1-{r_h\over r}\right)
\ ,
\end{array}
\ee
where $C$ is a constant.
Note also that the limit $q\to 0$ yields a solution of the homogeneous
(purely Schwarzschild) equations ($\phi_0^{(1)}=\phi_0^{(0)}$ and
$\phi_2^{(1)}=0$).
\par
The exchange of energy between ingoing and outgoing waves becomes
more apparent by adding the homogeneous parts (\ref{homo}) and
reshuffling the constants $A$, $B$ and $C$ so as to write the
general solution as
\be
\!\!\!\!\!\!
\begin{array}{l}
\phi_0=A\,\strut\displaystyle{e^{+i\,\sigma}\over r-r_h}\,\left[
1+\tilde x\,{B-A\over A}\,\ln\left(1-{r_h\over r}\right)\right]
Y_l^m
\\
\\
\phi_2=B\,\strut\displaystyle{e^{-i\,\sigma}\over r}\,\left[
1+\tilde x\,{B-A\over B}\,\ln\left(1-{r_h\over r}\right)\right]
Y_l^{-m}
\ ,
\end{array}
\label{sol}
\ee
where $\tilde x\equiv q\,x>0$ is an effective coupling constant between
the EM field and the dilaton, and the constants $A=A_{lm}(\omega)$ and
$B=B_{lm}(\omega)$ must be determined from the initial conditions.
There is obviously no initial condition (choice of $A$ and $B$)
which can produce just ingoing ($\phi_0$) or just outgoing
($\phi_2$) waves.
\subsection{Darkened stars}
Let us denote by $r_e>r_h$ the outer radius of the central star
and consider the case $A=0$, corresponding to a purely outgoing mode
$\phi_2$ in Schwarschild ($\tilde x=0$).
The ratio between the energy of such an EM wave ($\sim\phi_2^2$) at the
point of emission $r=r_e$ and at the point of observation $r=r_o$
would be
\be
\Gamma_{\rm Schw}\simeq
\left[{\phi_2^{(0)}(r_e)\over\phi_2^{(0)}(r_o)}\right]^2
\sim \left({r_o\over r_e}\right)^2
\ .
\ee
The same quantity in JNW (evaluated to leading order in $\tilde x$)
is obtained from the solution in Eq.~(\ref{sol}),
\be
\Gamma_{\rm JNW}\sim
\left({r_e\over r_o}\right)^2\,
{1+2\,\tilde x\,(1-{1/B})\,\ln\left(1-{r_h\over r_e}\right)\over
1+2\,\tilde x\,(1-{1/B})\,\ln\left(1-{r_h\over r_o}\right)}
\ ,
\ee
where $B$ can be determined, e.g.~from the energy $E_o$
at the point of observation,
\be
B^2\sim r_o^2\,E_o
\ .
\ee
On assuming that $r_h\ll r_e\ll r_o$ (a safe assumption to
protect the star from the central singularity), one finds
\be
\Gamma_{\rm JNW}\sim
\Gamma_{\rm Schw}\,\left[1
-2\,\tilde x\,\left(1-{1\over r_o\,\sqrt{E_o}}\right)\,
{r_h\over r_e}\right]
\ ,
\ee
which shows a peculiar dependence on the energy of the wave.
This dependence is not seen in the Schwarzschild case.
In particular, since $\tilde x>0$ and the term in round brackets
is positive for realistic cases, the flux is weaker than in the
Schwarzschild vacuum.
\par
With respect to what was found for the Kerr-Newman-dilaton black
hole \cite{knd2}, let us note that the final result does not show a
frequency dependence.
This can be understood since the static case we are now
considering has no typical frequency, whereas the rotating case
has intrinsic angular velocity.
\section{Freely falling source}
\label{free}
A possible source of electromagnetic radiation from a dilatonic
black hole is the radiation emitted by a charged particle falling
freely into the black hole.
In fact, the curved background distorts the EM field of the falling
charge and causes the latter to release energy as though it were
being accelerated in a Lorentz frame.
\par
For practical purposes, it is more convenient to regard the falling
charge $Q$ as distorting the background EM field $F_{\mu\nu}^{(0)}$
of the black hole, and expand the exact $F_{\mu\nu}$ to linear order
in $Q$ as
\be
F_{\mu\nu}=F_{\mu\nu}^{(0)}+Q\,F_{\mu\nu}^{(1)}
\ .
\ee
Since in our case $F_{\mu\nu}^{(0)}=0$, the perturbed EM tensor
elements $F_{\mu\nu}^{(1)}$ are determined by the Maxwell-dilaton
equation
\be
\partial_\nu\left(\sqrt{-g}\,e^{-q\,\Phi}\,F^{(1)\,\mu\nu}\right)
=4\,\pi\,\sqrt{-g}\,j^\mu \ ,
\label{max}
\ee
where $Q\,j^\mu$ is the current associated with the falling charge.
The metric tensor elements are given in Eq.~(2) and, in order to solve
the set of Eqs.~(\ref{max}), the (antisymmetric) perturbations
$F_{\mu\nu}^{(1)}$ are conveniently expanded in tensor harmonics
\cite{zerilli}
\be
F_{\mu\nu}^{(1)}=
\left[
{\begin{array}{cccc}
0  & f_{01}\,Y_l^m
& f_{02}\,\partial_\theta Y_l^m
& -f_{02}\,\sin(\theta)\,\partial_\phi Y_l^m
\\
[2ex]
* &  0
& f_{12}\,\partial_\theta Y_l^m
& f_{12}\,\partial_\phi Y_l^m
\\
[2ex]
* & * & 0 & 0
\\
[2ex]
* & * & 0 & 0
\end{array}}
\right]
\ ,
\label{tens}
\ee
where the $f_{ij}$'s depend only on $r$ and $t$ (and carry integer
indices $l$ and $m$ which we omit), the $Y_l^m$'s are
again the normalized spherical harmonics, and the asterisks ($*$)
denote the negative of the corresponding transposed tensor element.
The field tensor given in Eq.~(\ref{tens}) is for electric multipoles.
Further, using the field equations all of the tensor elements
$F_{\mu\nu}^{(1)}$ can be put in terms of a single element,
say $f_{12}$.
\par
The function $f_{12}$ is assumed to be of the form
\be
f_{12}(r,t)=\hat{f}_{12}(r)\,\exp(i\,\omega\,t)
\ .
\ee
Upon substitution of this expression into the equation for
$f_{12}$ and expanding to first order in $\tilde x\equiv q\,x$,
the function $\hat{f}_{12}$ satisfies the equation
\begin{subequations}
\begin{widetext}
\be
&&{d^2\hat{f}_{12}\over{dr^2}}
+{r_h\,(3+2\,\tilde x)\over{r\,(r-r_h)}}\,{d\hat{f}_{12}\over{dr}}
+\left[{\omega^2\,r^2\over (r-r_h)^2}
-{r_h\,(2\,r-3\,r_h)\over{r^2\,(r-r_h)^2}}\,(1+2\,\tilde x)
-{l\,(l+1)\,\over{r\,(r-r_h)}}\right]
\,\hat{f}_{12}
=S(r)
\ ,
\label{flm}
\ee
\end{widetext}
where the source term is given by
\be
S(r)=-{e^{i\,\omega\,T(r)}\over (r-r_h)^2}\,
\left[1-2\,\tilde x\,\ln\left(1-{r_h\over r}\right)\right]
\ ,
\label{source}
\ee
and
\be
T(r)=
r_h\left[2\,{r^{1/2}\over r_h^{1/2}}
-{2\,r^{3/2}\over 3\,r_h^{3/2}}
+\ln{(r/r_h)^{1/2}-1\over (r/r_h)^{1/2}+1}\right]
\ee
\end{subequations}
is the time (to order $\tilde{x}^2$) for the particle to fall
from $\infty$ to the point $r$.
\subsection{Green's function method}
Eq.~(\ref{flm}) is a typical Sturm-Liouville problem which can be
solved with the aid of the Green's function $G(r,r')$.
The latter is defined, for $r\not=r'$, as a solution of the homogenous
equation obtained from Eq.~(\ref{flm}) by setting $S(r)=0$ and, for
$r=r'$, by the boundary condition $G(r,r)=1$.
In terms of $G$ the solution of Eq.~(\ref{flm}) is then given by
\be
\hat{f}_{12}(r)=\int G(r,r')\,S(r')\,dr'
\ ,
\ee
where $S(r)$ is the source term (\ref{source})
\par
The homogenous equation which determines $G$ is of the form
\be
A_2\,G''+A_1\,G'+A_0\,G=0
\ ,
\label{homo12}
\ee
where the coefficients $A_i$ are determined by comparison with
Eq.~(\ref{flm}) and a prime denotes the partial derivative with
respect to the first argument of $G=G(r,r')$.
\par
Before attempting to solve Eq.~(\ref{homo12}) numerically, it is
interesting to estimate analytically the effect of the dilaton at
large distance from the singularity.
We fist expand the coefficients $A_i$'s in powers of $1/r$ and keep
only next-to-leading terms.
Then we expand the Green's function in powers of $\tilde x$ and write
$G=G_0+\tilde x\,G_1$.
To order $\tilde x^0$, Eq.~(\ref{homo12}) then yields
\begin{subequations}
\be
\hat L\,G_0&\equiv&
\left(1+3{r_h\over r}\right)G_0''
-3{r_h\over r^2}\,G_0'+\omega^2\left(1+{5\,r_h\over 3\,r}\right)G_0
\nonumber
\\
&=&0
\label{x0}
\ee
and, to order $\tilde x^1$,
\be
\hat L\,G_1
=\omega^2\,{5\,r_h\over 3\,r}\,G_0
\equiv S_G
\ .
\label{x1}
\ee
\end{subequations}
One thus finds that $G_1$ satisfies the same equation as $G_0$ plus
a source term determined by $G_0$ which does not vanish too rapidly
at large $r$.
One can therefore expect significant corrections with respect
to the Schwarzschild case ($\tilde x=0$) at relatively large distances
from the central singularity.
\par
It is not necessary to solve Eq.~(\ref{x1}) explicitly.
One can instead include the right hand side of Eq.~(\ref{x1}) in the
source $S$ and write the solution to Eq.~(\ref{flm}) simply in terms
of $G_0$ as
\be
\hat f_{12}(r)&=&\int G_0(r,r')\,S_0(r')\,dr'
+\tilde x\int G_0(r,r')\,S_1(r')\,dr'
\nonumber
\\
&&+\tilde x\,\int G_1(r,r')\,\left[S_0(r')+ S_G(r')\right]\,dr'
\nonumber
\\
&=& (1+\tilde x)\int G_0(r,r')\,S_0(r')\,dr'
\nonumber
\\
&&+ \tilde x \int G_0(r,r')\,\left[S_1(r') + S_G(r')\right]\,dr'
\ ,
\ee
in which we also expanded $S=S_0+\tilde x\,S_1$ and retained
terms to order $\tilde x^1$.
\par
Unfortunately Eq.~(\ref{x0}) is already rather complicated, and
the terms of higher order in $1/r$ which we dropped make the problem
intractable analytically.
Hence, we proceed now to solve Eq.~(\ref{flm}) numerically and
discuss the energy distribution.
\subsection{Numerical results}
\begin{figure}
\centering
\raisebox{3cm}{${1\over Q^2}\,{dE\over d\omega}$}
\epsfxsize=3in
\epsfbox{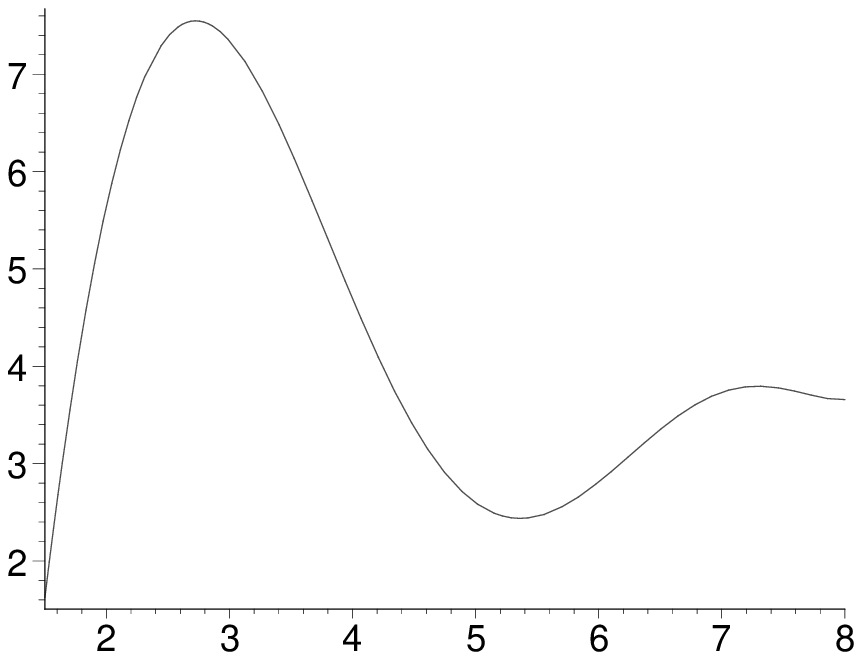}
\\
{\hspace{6cm}${\omega\,r_h}$}
\caption{Energy distribution for a particle of unit charge falling
into a four-dimensional black hole of mass $M$.}
\label{Edist}
\end{figure}

Once the function $\hat f_{12}$ has been determined, the energy
distribution corresponding to the angular mode $l$ and averaged
over the complete solid angle is given by
\be
\expecl{\frac{dE}{d\omega}}
={\frac{l\,(l+1)}{2\,\pi}}\,\hat{f}_{12}^{\ }\,\hat{f}^*_{12}
\ .
\label{dEdw}
\ee
The energy distribution (\ref{dEdw}) is plotted as a solid line
in Fig.~\ref{Edist} for unit charge $Q$ and black hole mass $M$
($=r_h/2$) and with the angular mode $l=1$.
The energy of the radiation is
\be
\Delta E\simeq C_{(4)}\,\frac{Q^2}{M}
\ ,
\label{En}
\ee
where $C_{(4)}$ is the area under the curve.
The secondary peak at higher frequencies in this distribution
distinguishes the dilatonic black hole from the Schwarzschild black hole,
which has only one peak (see the $2^{\rm nd}$ Reference of
\cite{zerilli}).
This suggests an interference between the dilatonic and tensor
contributions to the electromagnetic field components.
The secondary peak has an amplitude which is an appreciable fraction
of the higher peak at $\omega \approx 2.7$.
Such a peak should be relatively easy to detect.
The observation of such an interference term would be evidence for
the existence of a scalar component of gravity.
\section{Conclusions}
\label{conc}
We have analyzed the propagation of EM waves in the background
of the Janis-Newman-Winicour (JNW) solution.
To obtain manageable expressions describing the relevant physical
quantities, e.g.~the energy of the emitted waves and the intensity
as a function of frequency, we have had to expand the metric tensor
elements in terms of a small parameter $\tilde{x}$ which measures
the deviation of the JNW solution from the Schwarzschild solution
at large distance from the singularity.
This is a reasonable thing to do because the effect of a scalar
component of gravity on the aforementioned quantities is most likely
small.
We have applied the resulting expansions to two cases of physical
interest.
\par
The first case is that of a star with a dilatonic field.
We have obtained approximate analytic expressions for the ingoing and
outgoing wave functions in terms of the small expansion parameter,
$\tilde{x}$.
From these expressions we can calculate the luminosity of the object
emitting the radiation and we find that its luminosity is reduced by the
dilaton background.
Such a reduction would affect the luminosity-to-distance relations
that are used to determine the distance of astrophysical objects.
\par
In the second case we have described the emission from a point-like
charged particle freely falling in the chosen background.
In this case the JNW metric is expanded in terms of the small parameter
$\tilde{x}$ in order to obtain a simple form for the second order
differential equation whose solution is (the Fourier transform of) the
wave amplitude.
We have checked that setting the parameter $\tilde{x}$ to zero gives the
same differential equation as was obtained in \cite{zerilli} for the
Schwarzschild case.
For nonzero $\tilde{x}$ the frequency distribution curve $dE/d\omega$
differs significantly from that obtained for the Schwarzschild metric.
Models of the radiation from compact astrophysical sources incorporating
this effect would provide a test of the existence of a scalar component
of gravity.
\begin{acknowledgments}
This research was supported in part by DOE Grant No. DE-FG02-96ER40967.
\end{acknowledgments}

\begin{thebibliography}{99}
%
%
%
\bibitem{will}
C.M.~Will, {\it Theory and experiment in gravitational physics},
2nd ed., Cambridge University Press 1993;
Living Rev. Rel. {\bf 4} (2001) 4.
%
\bibitem{jnw}
A.I.~Janis, E.T.~Newman and J.~Winicour, Phys. Rev. Lett.
{\bf 20} (1968) 878.
%
\bibitem{bozza}
V.~Bozza, Phys. Rev. D {\bf 66} (2002) 103001.
%
\bibitem{chandra}
S.~Chandrasekhar, {\em The Mathematical Theory of Black Holes}
(Oxford University Press, Oxford, 1983).
%
\bibitem{knd0}
R.~Casadio, B.~Harms, Y.~Leblanc and P.H.~Cox, Phys. Rev. D
{\bf 55}, 814 (1997).
%
\bibitem{knd2}
R.~Casadio and B.~Harms, Phys. Rev. D {\bf 60}, 104017 (1999).
%
\bibitem{NSreview}
T.~Harada, H.~Iguchi and K.~Nakao, Prog. Theor. Phys. {\bf 107},
449 (2002).
%
\bibitem{finelli}
F.~Finelli and A.~Gruppuso, in preparation.
%
\bibitem{knd}
R.~Casadio, B.~Harms, Y.~Leblanc and P.H.~Cox, Phys. Rev. D
{\bf 56}, 4948 (1997).
%
\bibitem{zerilli}
F. Zerilli, Phys. Rev. D {\bf 9}, 860 (1974); M. Johnston, R.
Ruffini, F. Zerilli, Phys. Lett. {\bf B49}, 185 (1974).
%
\end{thebibliography}
\end{document}